\documentclass[prd,aps,twocolumn,a4paper,showkeys,nofootinbib]{revtex4-1}
\usepackage{amsmath}
\usepackage{amsfonts}
\usepackage{amssymb}	
\usepackage{graphicx}
\usepackage{bm}
\usepackage{color}

\def\p{\partial}

\def\GMc2{G M_{\odot} c^{-2}}

\def\vareps{\varepsilon}

\def\lm{{\ell m}}

\def\lm{{\ell m}}

\def\de{\partial}
\def\lm{{\ell m}}

\def\ii{{\rm i}}

\def\F{{\cal F}}

\def\TEOBResumS{\texttt{TEOBResumS}}
\def\TEOBResumROM{\texttt{TEOBResum\_ROM}}
\def\SEOBNRvq{{\texttt{SEOBNRv4}}}
\def\SEOBNRvqT{{\texttt{SEOBNRv4T}}}
\def\CC{{C\nolinebreak[4]\hspace{-.05em}\raisebox{.4ex}{\tiny\bf ++}}}

\begin{document}
\title{Rush the inspiral: efficient Effective One Body time-domain gravitational waveforms}

\author{Alessandro \surname{Nagar}${}^{1,2,3}$}
\author{Piero \surname{Rettegno}${}^{2,4}$}
\affiliation{${}^1$Centro Fermi - Museo Storico della Fisica e Centro Studi e Ricerche Enrico Fermi, 00184 Rome, Italy}
\affiliation{${}^2$INFN Sezione di Torino, Via P. Giuria 1, 10125 Torino, Italy}
\affiliation{${}^3$Institut des Hautes Etudes Scientifiques, 91440 Bures-sur-Yvette, France}
\affiliation{${}^{4}$ Dipartimento di Fisica, Universit\`a di Torino, via P. Giuria 1, 10125 Torino, Italy}

\begin{abstract}
  Computationally efficient waveforms are of central importance for gravitational wave
  data analysis of inspiralling and coalescing compact binaries. We show that the
  post-adiabatic (PA) approximation to the effective-one-body (EOB) description of
  the binary dynamics, when pushed to high order, allows one to accurately and
  efficiently compute the waveform of coalescing binary neutron stars (BNSs) or
  black holes (BBHs) up to a few orbits before merger.  This is accomplished 
  bypassing the usual need of numerically solving the relative EOB dynamics described by
  a set of ordinary differential equations (ODEs). Under the assumption 
  that radiation reaction is small, Hamilton's equations for the momenta can be solved
  {\it analytically} for given values of the relative separation. Time and orbital phase are then
  recovered by simple numerical quadratures. For the least adiabatic BBH case, equal-mass,
  quasi-extremal spins anti-aligned with the orbital angular momentum, 6PA/8PA orders
  are able to generate waveforms that accumulate less than $10^{-3}$~rad of phase difference
  with respect to the complete EOB ones up to $\sim 3$~orbits before merger.
  Analogous results  hold for BNSs. The PA waveform generation is extremely 
  efficient: for a standard BNS system from 10 Hz, a nonoptimized {\tt Matlab} 
  implementation of the \TEOBResumS~EOB model in the PA approximation 
  is almost 100 times faster ($\sim 0.09$~s) than the corresponding 
  \CC{} code  based on a standard ODE solver. Once optimized further, 
  our approach will allow us to (i) avoid the use of the fast, but often inaccurate, 
  post-Newtonian inspiral waveforms, drastically reducing the impact of systematics 
  due to inspiral waveform modelling, and (ii) alleviate the need of constructing EOB 
  waveform surrogates to be used in parameter estimation codes.
 \end{abstract}

\maketitle

\section{Introduction}
Analytical waveform models informed by (or calibrated to) numerical relativity (NR)
simulations are essential for the analysis of gravitational wave (GW)
events~\cite{Abbott:2016blz,Abbott:2016nmj,TheLIGOScientific:2016pea,Abbott:2017vtc,TheLIGOScientific:2017qsa,Abbott:2017gyy,Abbott:2017oio}.
The effective-one-body (EOB) approach to the general relativistic two body
problem~\cite{Buonanno:1998gg,Buonanno:2000ef,Damour:2001tu,Damour:2000we} 
is currently the only available analytical tool that reliably describes both the 
dynamics and the gravitational waveform through inspiral, merger and ringdown for binary black holes (BBHs)~\cite{Bohe:2016gbl,Nagar:2017jdw,Cotesta:2018fcv},
and up to merger for binary neutron stars (BNSs)~\cite{Dietrich:2017feu}. The analytical model is
crucially improved in the late-inspiral, strong-field, fast-velocity regime
by numerical relativity (NR) information, which allows one to properly represent
the merger and ringdown part of the waveform~\cite{Damour:2014yha,Bohe:2016gbl,Nagar:2017jdw}.
The synergy between EOB and NR creates EOBNR models, whose more recent
avatars are~\SEOBNRvq/\SEOBNRvqT~\cite{Bohe:2016gbl,Cotesta:2018fcv,Hinderer:2016eia,Steinhoff:2016rfi} and \TEOBResumS~\cite{Nagar:2017jdw,Dietrich:2018uni},
which describe nonprecessing binaries
and {\tt SEOBNRv3}~\cite{Babak:2016tgq}, which incorporates precession for BBHs.
However, it has to be understood that, though the framework is analytical,
the Hamiltonian equations of motion have to be solved {\it numerically}
with standard techniques. The computational cost of computing an EOB waveform
is then mostly due to the solution of Hamilton's equations: the longer the waveform,
the more expensive is its generation. For low-mass binaries, with long inspirals
within the sensitivity band of the detector, the computational cost of generating
an EOB waveform with~\TEOBResumS{}~\cite{Nagar:2018zoe} is $\sim$~ a few 
seconds\footnote{It is of the order of hours with~\SEOBNRvq~\cite{Bohe:2016gbl}.} 
is such that the model cannot be directly used for data analysis
purposes~(see, however, Refs.~\cite{Pankow:2015cra,Lange:2018pyp}; by contrast, it was possible
to use {\tt SEOBNRv3} explicitly on high-mass binaries~\cite{Abbott:2017vtc,Abbott:2016izl}). 
This prompted several recent efforts to optimize EOB codes~\cite{Devine:2016ovp,Knowles:2018hqq} 
or to compute {\it surrogate EOB waveforms} based on reduced-order modeling (ROM)
techniques~\cite{Field:2013cfa,Purrer:2014fza,Purrer:2015tud,Galley:2016mvy,Lackey:2016krb,Bohe:2016gbl}.
Building such surrogates is currently an obliged path to use EOB models
in standard data analysis pipelines. In addition, closed-form frequency-domain
phenomenological (Phenom) waveform models offer a valid alternative~\cite{London:2017bcn,Khan:2015jqa}.
These models are obtained by first joining together EOB-based inspirals with
NR simulations describing the last orbits through merger and ringdown and 
then building suitable interpolating fits all over the parameter space.
Though both EOB (surrogate) and Phenom proved comparatively good from the
BBH waveform generation point of view, EOB models are physically richer 
because of the built-in description of the relative dynamics. As a drawback, the
construction of these models is time consuming and not
very flexible. For example, if the original model is changed, the surrogate has 
to be rebuilt. Thus, an intermediate step (i.e., construction of a surrogate) is always
needed before new theoretical ideas can be tested on GW experimental data.
The detection and subsequent analysis of GW170817~\cite{TheLIGOScientific:2017qsa}
illustrated that the current status is far from optimal: the most developed
models with tidal interactions could not be immediately used on the data because
of their computational inefficiency; tidal EOB surrogates were not available except 
for the special nonspinning case~\cite{Lackey:2016krb}; PN-based and/or
Phenom-like tidal models~\cite{Dietrich:2017aum,Dietrich:2018uni} were available,
but they might be plagued by systematic effects that have to be
understood thoroughly~\cite{Dietrich:2018uni}, as they may result in biases
in the measured parameters. In this paper we follow a different route. 
Focusing on spin-aligned (nonprecessing) binaries, we use the {\it post-adiabatic} 
approximation to the EOB inspiral (using the \TEOBResumS~model~\cite{Nagar:2018zoe})
to obtain the gauge-invariant dynamics {\it analytically}, without the need of
solving ODEs. Through two additional quadratures, we then  obtain computationally
inexpensive, though robust and accurate, EOB inspiral waveforms up 
to a few orbits before merger. This is expected to tame, if not completely 
remove, most of the problems mentioned above.

\section{The Post--Adiabatic approximation to EOB inspiral}
The circularized binary dynamics of two objects with masses $(m_A,m_B)$
evolves quasi-adiabatically under the action of a gravitational
radiation reaction. Long ago, Ref.~\cite{Damour:2000zb}
constructed {\it resummed} inspiral waveforms
based on the adiabatic approximation (i.e., the dynamics
is represented by a sequence of circular orbits). When compared
to state-of-the-art 3PN EOB dynamics, they proved to be more
reliable and robust than the corresponding PN approximants.
However, the system inspirals inward because of the presence
of a small, but non-negligible, radial momentum $P_R$, and
the dynamics becomes progressively less adiabatic
as the merger is approached. The need of analytically
computing post-adiabatic (PA, linear in $P_R$)
corrections was pointed out as early as in Refs.~\cite{Buonanno:1998gg,Buonanno:2000ef},
so to provide low-eccentricity initial data to the EOB dynamics
when it is started at relatively close ($R=15M$) separations.
When high-accuracy, low-eccentricity NR data came
into play~\cite{Boyle:2007ft}, EOB/NR comparisons prompted
the need of {\it post-post-adiabatic} (2PA) initial
data~\cite{Damour:2007yf,Damour:2012ky}, 
so as to reduce the EOB eccentricity below the NR one.
The PA approximation to EOB dynamics is built as follows.
We use phase-space dimensionless variables $(r,p_{r_*},\varphi,p_\varphi)$, 
related to the physical ones by $r = R/GM$, $p_{r_*}= P_{R_*} / \mu$,
$p_\varphi = P_\varphi / (\mu GM)$, and $t = T/(GM)$, where $\mu=m_A m_B/M$,
with $M=m_A+m_B$. The radial momentum $p_{r_*}$ is defined as $p_{r_*} = (A/B)^{1/2} \ p_r$, in which $A$ and $B$ are the EOB potentials. Following Ref.~\cite{Damour:2014sva},
the EOB Hamiltonian is $\hat{H}_{\rm EOB}=\left(\sqrt{1+2\nu(\hat{H}_{\rm eff}-1)}\right)/\nu$,
where $\nu=\mu/M$ and $\hat{H}_{\rm eff}=\tilde{G}p_\varphi + \hat{H}^{\rm orb}_{\rm eff}$,
with $\tilde{G}p_\varphi$ being the spin-orbit sector, while $\hat{H}^{\rm orb}_{\rm eff}$
looks formally like the orbital (nonspinning) effective Hamiltonian,
though it actually incorporates in special resummed form spin-spin effects
through the use of the concept of centrifugal radius $r_c$~\cite{Damour:2014sva}. Explicitly, it reads $\hat{H}_{\rm eff}^{\rm orb} = \sqrt{p_{r_*}^2+A\big(1+p_\varphi^2 / r_c^2 +z_3\ p_{r_*}^4 / r_c^2\big)}$, with $z_3 = 2\nu(4-3\nu)$.
In addition, $\tilde{G} \equiv G_S \hat{S} + G_{S_*} \hat{S}_*$, where $\hat{S}\equiv (S_A+S_B)/M^2$,
$\hat{S}_*\equiv [(m_B/m_A)S_A + (m_A/m_B)S_B]/M^2$, in which $(S_A,S_B)$ are the individual spins,
and $(G_S,G_{S_*})$ are the gyrogravitomagnetic functions considered at
(next-to)$^2$-leading order. The spin gauge is fixed so that they only
depend on $(r,p_{r_*})$~\cite{Damour:2007nc,Nagar:2011fx,Damour:2014sva}.
An effective (next-to)$^3$-leading-order parameter
$c_3$~\cite{Damour:2014sva,Nagar:2015xqa} is included in $(G_S,G_{S_*})$
and informed by NR simulations as in Ref.~\cite{Nagar:2018zoe}.
The EOB Hamilton's equations for spin-aligned binaries that are usually solved numerically read
\begin{align}
\label{eq:dvarphidt}
\frac{d\varphi}{dt}=& \ \frac{1}{\nu \hat{H}_{\rm EOB} \hat{H}_{\rm eff}^{\rm orb} }\Big[A \frac{p_\varphi}{r_c^2}+\hat{H}_{\rm eff}^{\rm orb} \tilde{G}\Big],\\
\label{eq:drdt}
\frac{dr}{dt}=& \ \Big(\frac{A}{B} \Big)^{1/2} \frac{1}{\nu \hat{H}_{\rm EOB} \hat{H}_{\rm eff}^{\rm orb}} \times \nonumber \\
&\times\Big[ p_{r_*} \Big(1+2z_3 \frac{A}{r_c^2} p_{r_*}^2\Big) + \hat{H}_{\rm eff}^{\rm orb} p_\varphi \frac{\p \tilde{G}}{\p p_{r_*}} \Big],\\
\label{eq:dpphi}
\frac{dp_\varphi}{dt}=& \ \hat{\F}_\varphi,\\
\label{eq:dprdt}
\frac{dp_{r_*}}{dt}=& -\Big(\frac{A}{B} \Big)^{1/2} \frac{1}{2 \nu \hat{H}_{\rm EOB} \hat{H}_{\rm eff}^{\rm orb} } \Big[A'+ p_\varphi^2 \Big(\frac{A}{r_c^2}\Big)'+\nonumber \\
& \hspace{0.5cm}+ z_3 \ p_{r_*}^4\Big(\frac{A}{r_c^2}\Big)' +2 \hat{H}_{\rm eff}^{\rm orb} p_\varphi \tilde{G}' \Big],
\end{align}
where $(\cdot)^\prime \equiv \de_r(\cdot)$, and we fix $\hat{\F}_{r_*} = 0$ in Eq.~\eqref{eq:dprdt}. The $A$ function
incorporates an effective 5PN parameter $a_6^c(\nu)$ informed by
NR simulations~\cite{Nagar:2017jdw}. Tidal effects, as well as spin-induced quadrupole-moment
effects, are also included in the formalism~\cite{Damour:2009wj,Bini:2012gu,Damour:2012yf,Bernuzzi:2014owa,Bernuzzi:2014kca,Bernuzzi:2015rla,Hinderer:2016eia,Steinhoff:2016rfi,Nagar:2018zoe}. 
The adiabatic approximation assumes no radiation reaction, $\hat{\F}_\varphi = 0$,
so that $p_{r_*} = 0$ and $p_\varphi = j_0$,
obtained imposing $\de_r \hat{H}_{\rm EOB} = 0$ at a
given radius $r$ [see Eq.~(C1) of~\cite{Nagar:2018zoe}].
The PA approximation~\cite{Buonanno:2000ef}
assumes $\hat{\F}_\varphi$ to be small and then consistently calculates $p_{r_*}$,
combining Eqs.~\eqref{eq:drdt} and \eqref{eq:dpphi} as $dp_\varphi/dr=\hat{\F}_\varphi (dr/dt)^{-1}$.
At the 2PA level~\cite{Damour:2007yf}, one obtains the additional correction
to $p_\varphi$ as $p_{r_*}\neq 0$ using Eq.~\eqref{eq:drdt} and~\eqref{eq:dprdt}.
The procedure can then be iterated further.
At a formal level, one is assuming that, for each $r$,
$\hat{\F}_\varphi (r) =  \ \sum_{n = 0}^{\infty} \F_{2n+1}(r) \ \vareps^{2n+1}$,
where $\vareps$ is a formal bookkeeping parameter that is eventually set to $1$.
We can hence write the solution of the EOB equations of motion as a formal expansion
in powers of $\vareps$ as 
$p_\varphi^2(r) = \ j_0^2(r) \Big(1+ \sum_{n = 1}^{\infty} k_{2n}(r) \ \vareps^{2n}\Big)$ ,
and
$p_{r_*}(r) = \sum_{n = 0}^{\infty} \pi_{2n+1}(r) \ \vareps^{2n+1}$.
To finally obtain the corrections ($k_{2n},\pi_{2n+1})$ to the circular solution $(j_0,0)$,
one has to iteratively solve the two equations
$dp_{r_*}/dt = (dp_{r_*}/dr)(dr/dt)$ and $dp_\varphi/dt = (dp_\varphi/dr)(dr/dt)$,
where the radial derivatives of the momenta are replaced by their
series expansion above, and the time-derivatives are substituted
by Hamilton's equations.
We solve these equations alternatively power by power in $\vareps$,
obtaining the corrections to the angular and radial momentum
respectively. In general, every coefficient depends on the lower-order ones and their radial derivatives. This procedure can be
iterated as many times as one likes. We call $n$-post-adiabatic ($n$PA)
a quantity calculated up to $\vareps^n$. Computing the $n$PA approximation
is then straightforward, since only linear equations are involved, though
tedious at high order, as it involves many terms.
\begin{figure*}[t]
\center
\includegraphics[width=0.32\textwidth]{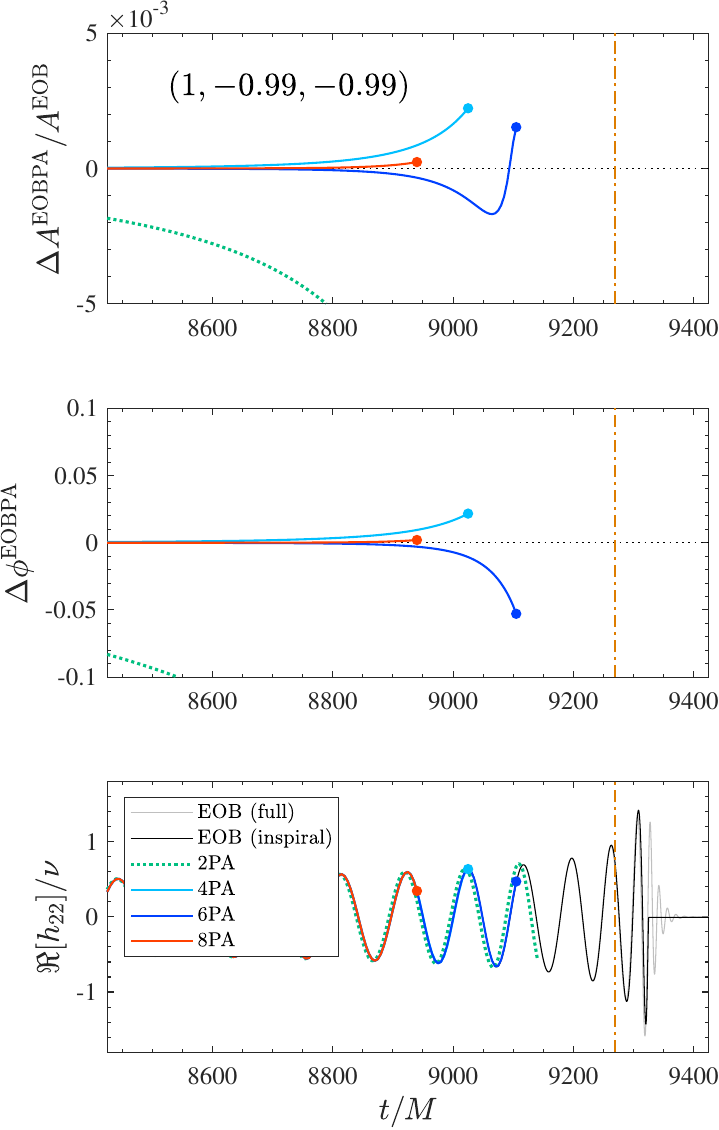}
\includegraphics[width=0.32\textwidth]{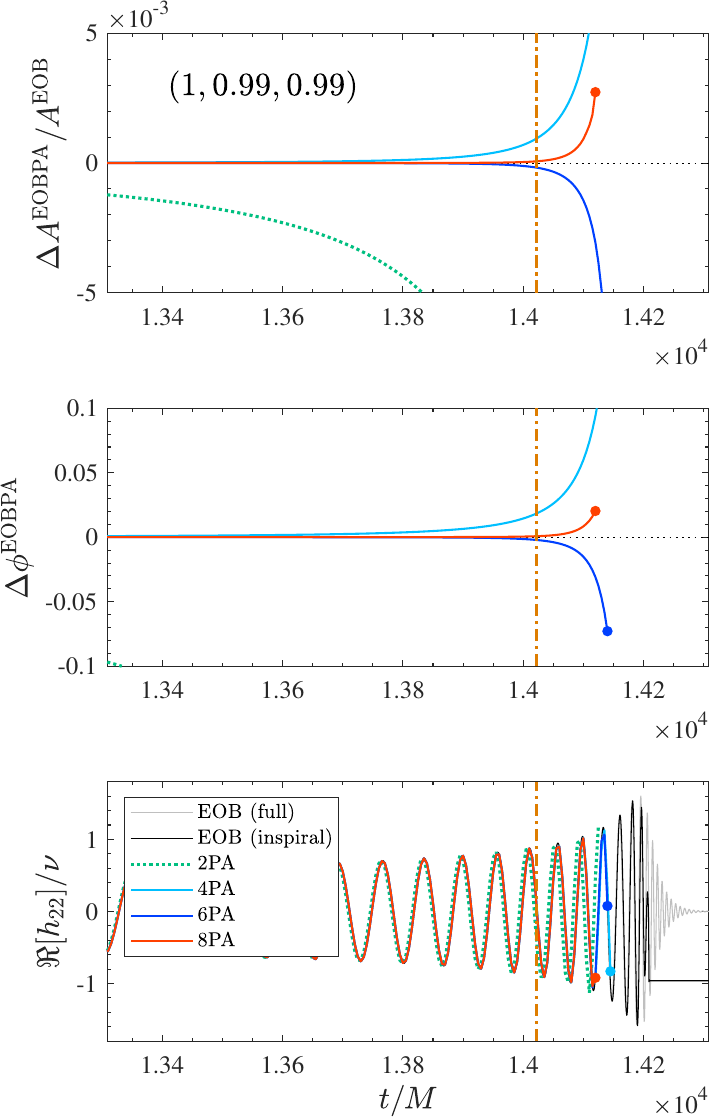}
\includegraphics[width=0.32\textwidth]{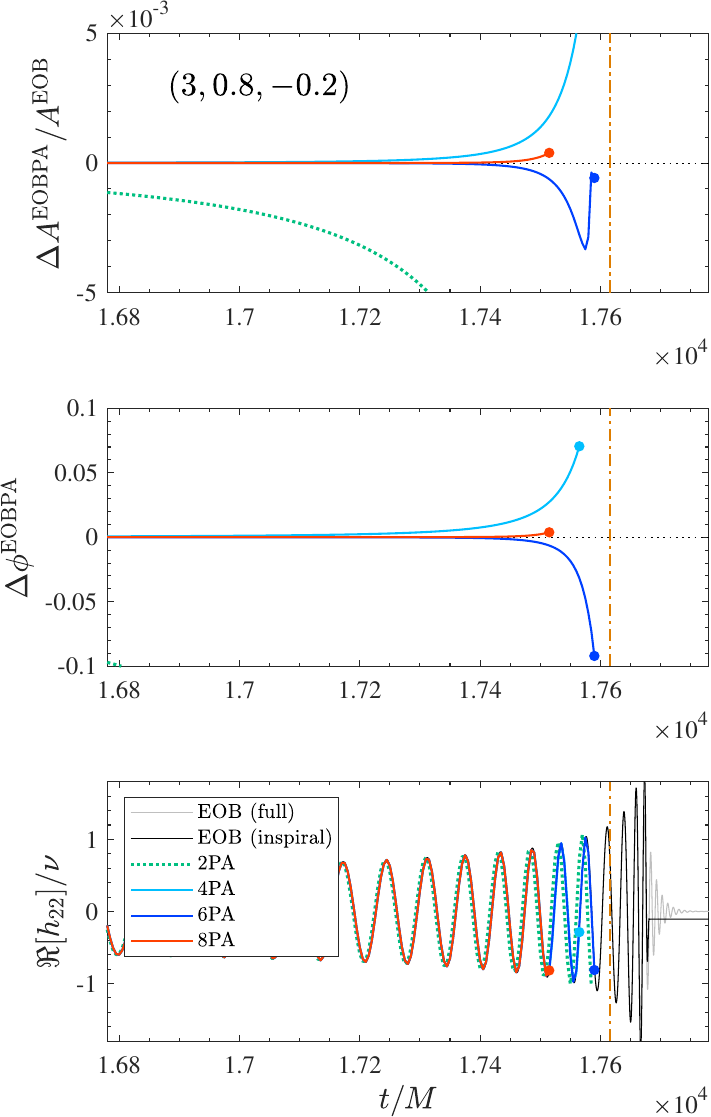}
\caption{\label{fig:wave_bbh} Waveform comparison, $\ell=m=2$ strain mode:
  EOB$_{\rm PA}$ inspiral (colors) versus EOB inspiral obtained solving the ODEs (black).
  The  light-gray curve also incorporates the EOB merger and ringdown.
  The orange vertical line marks the EOB LSO crossing for $(1,-0.99,-0.99)$ and $(3,+0.80,-0.20)$,
  while it corresponds to $r=6$-crossing for $(1,+0.99,+0.99)$.
  The filled markers highlight the end of the PA inspirals.
  The 4PA approximation
  already delivers an acceptable EOB/EOB$_{\rm PA}$ agreement for both phase,
  $\phi$, and amplitude, $A$. This is improved further by the successive
  PA approximations. At 8PA, the GW phase difference is $\lesssim 10^{-3}$~rad
  up to $\sim 3$ orbits before merger. }
\end{figure*}
\begin{figure}[t]
\center
\includegraphics[width=0.48\textwidth]{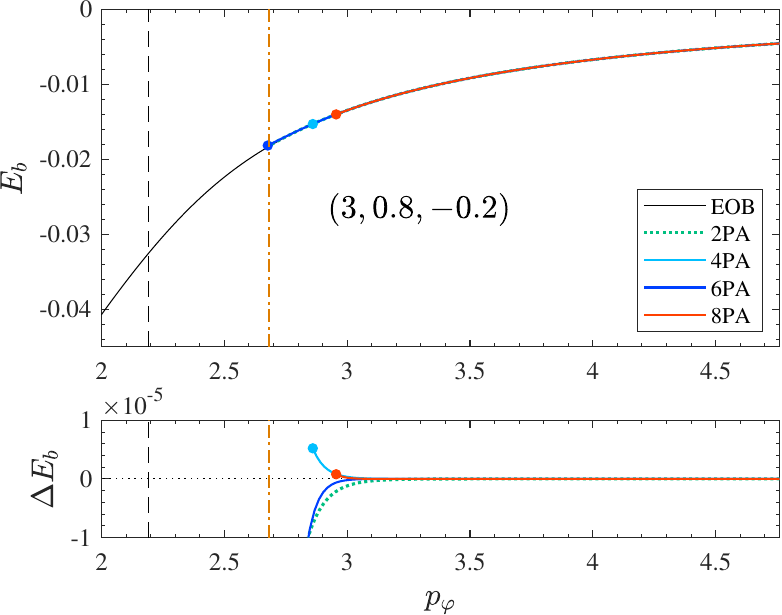}
\caption{\label{fig:E_j} Illustrative comparison between energies
  versus orbital angular momentum curves. The black vertical line
  marks the EOB merger, while the orange one marks the EOB-LSO crossing.
  The EOB/EOB$_{\rm PA}$ agreement is excellent up to the range
  of validity of each PA order (filled markers).}
\end{figure}
We can obtain the same results following a mathematically less rigorous but quicker path.
We can express the two equations above as
\begin{align}
\label{eq:pphi}
&\Big(\frac{A}{r_c^2}\Big)'p_\varphi^2 + 2 \hat{H}_{\rm eff}^{\rm orb} \Big(\frac{\p \tilde{G}}{\p r} + \frac{\p \tilde{G}}{\p p_{r_*}} \frac{d p_{r_*}}{d r}\Big) p_{\varphi} + \nonumber \\
& + A' + 2\Big(1+2 z_3 \frac{A}{r_c^2} p_{r_*}^2\Big)p_{r_*}\frac{d p_{r_*}}{dr} + z_3\Big(\frac{A}{r_c^2}\Big)' p_{r_*}^4 = 0, 
\end{align}
\begin{align}
\label{eq:prs}
&p_{r_*} = \hat{\F}_\varphi\left(\frac{d p_\varphi}{d r}\right)^{-1} \Big(\frac{A}{B}\Big)^{-1/2} \nu \hat{H}_{\rm EOB} \hat{H}_{\rm eff}^{\rm orb} \ \times \nonumber \\
& \hspace{1cm}\times \Big(1+2 z_3 \frac{A}{r_c^2} p_{r_*}^2 + p_\varphi \hat{H}_{\rm eff}^{\rm orb} \Big[\frac{1}{p_{r_*}} \frac{\p \tilde{G}}{\p p_{r_*}}\Big] \Big)^{-1}.
\end{align}
We treat the explicit $p_\varphi$ in the first equation and the $p_{r_*}$
in the left hand side of the second as the only unknown variables. All other
$(p_\varphi,p_{r_*})$'s that appear, also within $(\hat{H}_{\rm EOB},\hat{H}_{\rm eff}^{\rm orb},\tilde{G},\dots)$,
are kept at previously known order. This is easier to implement, as the same equations
must be solved at each order. Since we did not find any significant discrepancy with
the rigorous PA approximation, we only present here results with
Eqs.~\eqref{eq:pphi} and \eqref{eq:prs}. The $n$PA dynamics is then obtained via
a three-step procedure: (i) A radial grid is built 
between $r_{\rm max}$ and $r_{\rm min}$, with spacing $\Delta r$ 
chosen uniform for simplicity. (ii) For each grid point, $(p_\varphi,p_{r_*})$
are obtained from Eqs.~\eqref{eq:pphi} and \eqref{eq:prs}
at a given iteration order. (iii) The time $t$ and the orbital
phase $\varphi$ are recovered by quadratures
as $t = \int_{r_{\rm max}}^r dr (\partial_{p_r}\hat{H})^{-1}$ and
$\varphi =  \int_0^t dt \p_{p_\varphi}\hat{H} = \int_{r_{\rm max}}^r dr \p_{p_\varphi}\hat{H} (\partial_{p_r}\hat{H})^{-1}$.
This way, one obtains $(t,\varphi,p_{\varphi},p_{r_*})$ on a given $r$ grid.
Since the $r$ grid is evenly spaced, $t$ is not, with time steps becoming
progressively smaller as $r$ decreases. 
Finally, though the $(r_{\rm max},r_{\rm min})$ grid is built with $r_{\rm min}$
near to the EOB LSO, the physically meaningful range of the PA approximation
is only up to a given $r_{\rm inspl\text{-}end}>r_{\rm min}$
where $r_{\rm inspl\text{-}end}$ corresponds to the first inflection point of $p_{r_*}$ 
i.e., where it starts decreasing more slowly instead of keeping on accelerating,
as is the case for the complete EOB. This point depends on the PA order
and will be explicitly marked when discussing results below. The waveform is
built upon the PA dynamics $(t,\varphi,p_\varphi,r,p_{r_*})$ using the standard
EOB prescription~\cite{Damour:2011fu,Damour:2014sva,Nagar:2015xqa}.
The multipolar strain waveform $h_\lm$ is defined as
$h_+ -\ii h_\times ={\cal R}^{-1}\sum_{\ell,m}h_{\ell m} {}_{-2}Y_{\ell m}(\theta,\Phi)$,
where ${\cal R}$ is the distance from the source and  ${}_{-2}Y_{\ell m}(\theta,\Phi)$ is
the $s=-2$ spin-weighted spherical harmonics. Though we obtain all inspiral multipoles 
at once up to $\ell=8$, we only discuss $h_{22}=A e^{-{\rm i}\phi}$. We refer
to the PA approximated EOB as EOB$_{\rm PA}$.

\section{Results}
The quality of the EOB$_{\rm PA}$ model is assessed on a few relevant cases
involving spin-aligned BBHs and BNSs. Figure~\ref{fig:wave_bbh} illustrates
the phasing performance for three fiducial BBHs with $(q,\chi_A,\chi_B)$ equal
to $(1,-0.99,-0.99)$, $(1,+0.99,+0.99)$ and $(2,+0.8,-0.2)$.
The first configuration allows us to test the approximation in the most difficult
regime i.e., when the inspiral is less adiabatic due to the strong, attractive,
spin-orbit interaction exerted by the two spins anti-aligned with the orbital
angular momentum. The various~PA approximations
are contrasted with the corresponding time-domain
\TEOBResumS \ waveform~\cite{Nagar:2017jdw,Dietrich:2018uni}.
The top row shows the relative amplitude difference
$\Delta A^{\rm EOBPA}/A^{\rm EOB}\equiv (A^{\rm EOB}-A^{\rm PA})/A^{\rm EOB}$ and the middle panel shows the phase difference 
$\Delta \phi^{\rm EOBPA}\equiv \phi^{\rm EOB}-\phi^{\rm PA}$, while the bottom panel
depicts the real part of the waveform. The black line is the pure {\it inspiral} 
EOB waveform, without the NR-informed next-to-quasi-circular (NQC) correction 
parameters or the ringdown~\cite{Nagar:2017jdw}. For completeness, we also added the
 {\it full} EOB waveform with merger and ringdown (gray line). 
For this specific comparison, we did not iterate on NQC parameters~\cite{Nagar:2017jdw}. 
The plot focuses on the last few cycles of inspirals that began at initial separation $r_{\rm max}=20$. 
The time-evolution of $\TEOBResumS$ was initiated with 2PA initial data. 
To orient the reader, the orange vertical line marks the location of the 
adiabatic EOB last stable orbit
(LSO) for $(1,-0.99,-0.99)$ ($r^{\rm EOB}_{\rm LSO}=6.69$) and $(3,+0.80,-0.20)$
($r^{\rm EOB}_{\rm LSO}=4.30$), while it corresponds to
the $r_{\rm LSO}^{\rm Schwarzschild}=6$ crossing for $(1,+0.99,+0.99)$
(the adiabatic  \TEOBResumS~dynamics 
does not have an LSO when the spins are large and aligned~\cite{Balmelli:2015zsa}). 
The PA approximation converges very fast, and moving from 2PA to 4PA 
is already sufficient to obtain phase differences $<0.05$~rad $\sim 3$ orbits before merger.
When pushed to higher order (notably 8PA) the phase difference
is $\lesssim 10^{-3}$~rad up to $\sim 3$ orbits before merger. 
Note that we {\it did not} perform any additional phase or time 
alignment between the waveforms. 
The PA waveforms were obtained with resolution $\Delta r=0.1$
and $r_{\rm min}\sim (8.8,4.2,4.1)$ respectively, with $N^{\rm PA}_r=(112,159,159)$
points. We verified that, thanks to the crucial fact that we use a third-order
integration routine to recover $[\varphi(r),t(r)]$,  the waveform temporal
length is insensitive to the choice of resolution, which can be safely increased
up to $\Delta r=0.4$. For illustrative purposes in Fig.~\ref{fig:wave_bbh},
the original, sparse and nonuniform, temporal grid
corresponding to $(r_{\rm max},r_{\rm inspl\text{-}end})$ is interpolated
on a uniform grid with $\Delta t=0.5M$. The end of the 8PA-inspiral corresponds to $r_{\rm inspl\text{-}end}=(10.55,4.1,5.9)$.
Figure~\ref{fig:E_j} compares the binding energy per reduced mass,
$E_b=(E-M)/\mu$, where $E=\nu \hat{H}$ computed along the EOB
and EOB$_{\rm PA}$ dynamics. The agreement is excellent nearly up to the
LSO and, notably, within the expected uncertainty on this quantity
computed from NR simulations~\cite{Damour:2011fu,Nagar:2015xqa,Ossokine:2017dge}.
\begin{figure}[t]
\center
\includegraphics[width=0.48\textwidth]{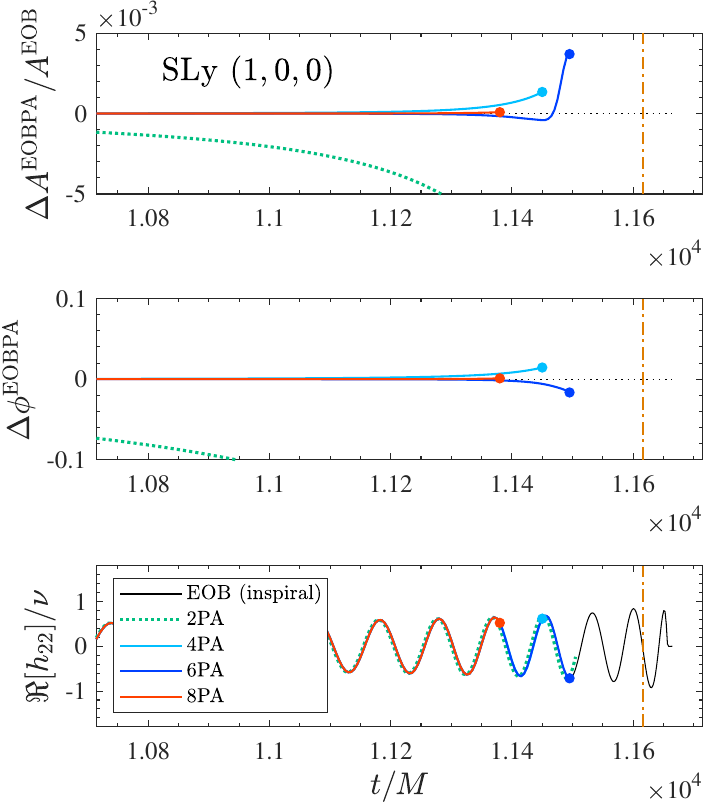}
\caption{\label{fig:BNS} Illustrative BNS case: SLy EOS, $m_A=m_B=1.35M_\odot$, $\chi_A=\chi_B=0$,
  quadrupolar tidal polarizabilities $\Lambda_A=\Lambda_B=392$. The orange vertical
  line is the EOB LSO crossing; the markers highlight the end of the PA inspirals.
  The 2PA $\Delta\phi^{\rm EOBPA}$ is outside the $y$ range and thus not shown.}
\end{figure}
 \begin{table}[t]
   \caption{\label{tab:performance}Illustrative performance comparison between
     the C\hspace{-.05em}\raisebox{.4ex}{\tiny\bf ++} implementation of \TEOBResumS\  with ODE solver and the corresponding
     8PA {\tt Matlab} implementation for the fiducial BNS system of Fig.~\ref{fig:BNS}.
     The sampling rate of the ODE is 4096~Hz, corresponding to $\Delta t\simeq 18M$.
     The EOB$_{\rm PA}$ waveform is obtained from $N_{r}^{\rm PA}$ radial points.
     The runs are done on a MacBook Pro, with Intel Core i7, 3.5GHz, 16GB RAM. }
   \begin{center}
 \begin{ruledtabular}
   \begin{tabular}{ccccc}
     $f_{0}$ [Hz] & $R_{\rm max}/M$ & $N_{r}^{\rm PA}$ & $\tau_{\rm run}^{\rm ODE}$ [s]  &$\tau_{\rm run}^{\rm PA}$ [s] \\
     \hline
     10   &  178.73  &  1143  &  6.12   &  0.09   \\
     20   &  112.73  &  702   &  1.17   &  0.08   \\
     30   &  86.029  &  524   &  0.49   &  0.065  \\ 

 \end{tabular}
 \end{ruledtabular}
 \end{center}
 \end{table}
Figure~\ref{fig:BNS} illustrates the similar behavior for a fiducial, equal-mass,
BNS system, with tidal polarizabilities $\Lambda_A=\Lambda_B=(2/3)k_2/C^5=392$, where $k_2$ is the quadrupolar relativistic Love
number~\cite{Damour:1983a,Hinderer:2007mb,Binnington:2009bb,Damour:2009vw,Hinderer:2009ca},
and $C$ is the star compactness. This picture is stable when changing equation
of state (EOS) and/or incorporating the spins.
For simplicity, the EOB$_{\rm PA}$ model we discuss here was implemented
in {\tt Matlab} without any optimization strategy. Table~\ref{tab:performance}
contrasts the performance of such {\tt Matlab} implementation with the \CC{} version
of \TEOBResumS~\cite{Nagar:2018zoe} for a few long inspirals.
The radial PA grid has $\Delta r=0.15M$, but, as before, the EOB$_{\rm PA}$ waveform
remains stable even with coarser grids up to $\Delta r=0.4$.
We note in passing that the 8PA running time is comparable to (and actually smaller than)
 the one provided by the \TEOBResumROM\ model for nonspinning BNSs 
of Ref.~\cite{Lackey:2016krb}.

\section{Conclusions}
We showed that the post-adiabatic analytic approximation to
the EOB dynamics, when pushed to high order, is a useful tool
to compute approximate, though reliable, EOB inspiral waveforms.
These EOB$_{\rm PA}$ waveforms well reproduce ($\Delta \phi\lesssim 0.001$~rad)
the non-approximated ones, obtained numerically solving Hamilton's
equations, up to $\sim 3$ orbits before merger, independently of the
(illustrative) binary configurations considered. For BNS inspirals,
Table~\ref{tab:performance} showed that even a largely 
nonoptimized {\tt Matlab} infrastructure can generate waveforms
whose computational cost is orders of magnitude smaller
than the dedicated \TEOBResumS\ \CC{} numerical code.
Though the PA approximation is not reliable in the last $\sim 3$ orbits,
nonetheless it can be used to start ODE-based EOB evolutions 
from a radius larger than $r_{\rm inspl\text{-}end}$ (e.g., twice as large,
to be conservative), so to reduce the computational cost 
of a complete EOB waveform.
For a BBH event, the \CC{} typically takes $\lesssim 16$~ms to ODE-evolve
the last seven orbits through merger and ringdown.
We expect that our approach, once properly implemented in
data-analysis pipelines, will allow one to avoid the use
of PN-based inspiral waveform models, so as to drastically
reduce the impact of systematics due to waveform modeling
on GW data analysis of long-inspiral coalescing compact
binaries, like GW170817. One could also combine
the PA approximation with the stationary phase approximation
to directly compute the Fourier-domain inspiral
phase~\cite{Buonanno:2009zt,Damour:2010zb}. The need of EOB waveform
surrogates might then be reduced. Due to its flexibility, the
${\rm EOB_{PA}}$ model could also be directly used to perform
tests of general relativity. Finally, we expect the PA approximation
to be useful for circularized precessing binaries, as well as for
the EOB inspiral of eccentric binaries~\cite{Damour:2004bz,Bini:2012ji,Hinderer:2017jcs,Cao:2017ndf},
although the impact of the radial part of the radiation reaction
needs to be properly evaluated.

A C implementation of ${\rm EOB_{PA}}$ is publicly available 
as a stand-alone code (see Ref.~\cite{Nagar:2018zoe}, Appendix~E).
Its performances are better than those of the {\tt Matlab}
one discussed here and will be detailed elsewhere.

\section*{Acknowledgments}
We thank S.~Akcay, S.~Bernuzzi, G.~Carullo, T.~Damour, W.~Del Pozzo, 
N.~Fornengo, P.~Schmidt and G.~Pratten for discussions.
We are grateful to C.~Chapman for inspiring ideas.


\bibliography{refs20191125.bib}

\end{document}